\documentclass[sigconf]{acmart}

\setcopyright{rightsretained}

\acmDOI{10.475/123_4}

\acmISBN{123-4567-24-567/08/06}

\acmConference[IUI]{ACM Conference on Intelligent User Interfaces}{March 2019}{Los Angeles, California USA}
\acmYear{2019}
\copyrightyear{2016}

\acmArticle{4}
\acmPrice{15.00}

\begin{document}
\title{Vulnerable to Misinformation? Verifi!}

\author{Alireza Karduni \\ Isaac Cho \\ Ryan Wesslen \\ Sashank Santhanam}
\orcid{1234-5678-9012}
\affiliation{%
  \institution{UNC-Charlotte, Department of Computer Science}
  \city{Charlotte}
  \state{North Carolina}
  \postcode{28213}
}
\email{{akarduni, icho1, rwesslen, ssantha1} @uncc.edu}

\author{Svitlana Volkova \\ Dustin L Arendt}
\orcid{1234-5678-9012}
\affiliation{%
  \institution{Pacific Northwest National Labarotory}
  \city{Richland}
  \state{Washington}
  \postcode{99354}
}
\email{{ svitlana.volkova, dustin.arendt} @pnnl.gov}

\author{Samira Shaikh \\ Wenwen Dou}
\orcid{1234-5678-9012}
\affiliation{%
  \institution{UNC-Charlotte, Department of Computer Science}
  \city{Charlotte}
  \state{North Carolina}
  \postcode{28213}
}
\email{{sshaikh2, wdou1}@uncc.edu}

\renewcommand{\shortauthors}{A. Karduni et al.}

\begin{abstract}
We present Verifi2, a visual analytic system to support the investigation of misinformation on social media. Various models and studies have emerged from multiple disciplines to detect or understand the effects of misinformation. However, there is still a lack of intuitive and accessible tools that help social media users distinguish misinformation from verified news. Verifi2 uses state-of-the-art computational methods to highlight linguistic, network, and image features that can distinguish suspicious news accounts. By exploring news on a source and document level in Verifi2, users can interact with the complex dimensions that characterize misinformation and contrast how real and suspicious news outlets differ on these dimensions. To evaluate Verifi2, we conduct interviews with experts in digital media, communications, education, and psychology who study misinformation. Our interviews highlight the complexity of the problem of combating misinformation and show promising potential for Verifi2 as an educational tool on misinformation. 
\end{abstract}

%
%

\copyrightyear{2019} 
\acmYear{2019} 
\setcopyright{licensedusgovmixed}
\acmConference[IUI '19]{24th International Conference on Intelligent User Interfaces}{March 17--20, 2019}{Marina del Ray, CA, USA}
\acmBooktitle{24th International Conference on Intelligent User Interfaces (IUI '19), March 17--20, 2019, Marina del Ray, CA, USA}
\acmPrice{15.00}
\acmDOI{10.1145/3301275.3302320}
\acmISBN{978-1-4503-6272-6/19/03}

\ccsdesc[500]{Human-centered computing~User interface design}
\ccsdesc[500]{Human-centered computing~Social media}
\ccsdesc[500]{Human-centered computing~Visual analytics}

\keywords{Misinformation, Social Media, Visual Analytics, Fake News}

\maketitle

\section{Introduction}

The rise of misinformation online has a far-reaching impact on the lives of individuals and on our society as a whole. Researchers and practitioners from multiple disciplines including political science, psychology and computer science are grappling with the effects of misinformation and devising means to combat it \cite{bourgonje2017clickbait, pennycook2017prior, shao2017spread, shu2017fake, SHEG:2016}. 

Although hardly a new phenomenon, both anecdotal evidence and systematic studies that emerged recently have demonstrated real consequences of misinformation on people's attitudes and actions \cite{PLOSONE2017, pizzagate}. There are multiple factors contributing to the widespread prevalence of misinformation online. On the content generation end, misleading or fabricated content can be created by actors with malicious intent. In addition, the spread of such misinformation can be aggravated by social bots. On the receiving end, cognitive biases, including confirmation bias, that humans exhibit and the echo chamber effect on social media platforms make individuals more susceptible to misinformation. 

Social media platforms have become a place for many to consume news. A Pew Research survey on ``Social Media Use in 2018'' \cite{social_media_use_in_2018} estimates a majority of Americans use Facebook (68\%) and YouTube (73\%). In fact, 88\% younger adults (between the age of 18 and 29) indicate that they visit any form of social media daily. However, despite being social media savvy, younger adults are alarmingly vulnerable when it comes to determining the quality of information online. According to the study by the Stanford History Education Group, the ability of younger generations ranging from middle school to college students that grew up with the internet and social media in judging the credibility of information online is bleak \cite{SHEG:2016}. The abundance of online information is a double-edged sword - \textit{``[it] can either make us smarter and better informed or more ignorant and narrow-minded''}. The authors of the report further emphasized that the outcome \textit{``depends on our awareness of the problem and our educational response to it''}.  

Addressing the problem of misinformation requires raising awareness of  
a complex and diverse set of factors not always visible in a piece of news itself, but in its contextual information. To this aim, we propose Verifi2, a visual analytic system that enables users to explore news in an informed way by presenting a variety of factors that contribute to its veracity, including the source's usage of language, social interactions, and topical focus. The design of Verifi2 is informed by 
recent studies on misinformation \cite{SHEG:2016, lazer2018science}, computational results on features contributing to the identification of misinformation on Twitter \cite{volkova2017separating}, and empirical results from user experiments \cite{karduni2018icwsm}.  

Our work makes the following contributions:
\begin{itemize}
    \item Verifi2 is one of the first visual interfaces that present features shown to separate real vs. suspicious news \cite{volkova2017separating}, thus raising awareness of  multiple features that can inform the evaluation of news account veracity.
    \item To help users better study the social interaction of different news sources, Verifi2 introduces a new social network visualization that simultaneously presents accounts' direct interactions and their topical similarities.
    \item Verifi2 leverages state-of-the-art computational methods  to support the investigation of misinformation by enabling the comparison of how real and suspicious news accounts differ on language use, social network connections, entity mentions, and image postings. 
    \item We provide usage scenarios illustrating how Verifi2 supports reasoning about misinformation on Twitter. We also provide feedback from experts in education/library science, psychology, and communication studies 
    as they discuss how they envision using such system within their work on battling misinformation and the issues they foresee.
\end{itemize}

\section{Production and Identification of Misinformation
}

Misinformation and its many related concepts (disinformation, malinformation, falsehoods, propaganda, etc.) have become a topic of great interest due to its effect on political and societal processes of countries around the world in recent years \cite{sanovich2017computational, miller2003tell, khaldarova2016fake}. However, it is hardly a new phenomenon. At various stages in history, societies have dealt with propaganda and misinformation coming from various sources including governments, influential individuals, mainstream news media, and more recently, social media platforms \cite{bakir2017fake}. It has been shown that misinformation indeed has some agenda setting power
with respect to partisan or non-partisan mainstream news accounts or can even change political outcomes \cite{vargo2017agenda, benkler2017study}. Misinformation is a broad and loosely defined concept and it consists of various types of news with different goals and intents. Some of the identified types of misinformation include satire, parody, fabricated news, propaganda, click-baits, and advertisement-driven news \cite{tandoc2017defining, tambini2017fake}. The common feature among these categories is that misinformation resembles the look and feel of real news in form and style but not in accuracy, legitimacy, and credibility \cite{lazer2018science, tandoc2017defining}. In addition to the concept of misinformation, there is a broader body of research on news bias and framing defined as ``to select some aspects of a perceived reality and make them more salient in a communicating text, in such a way as to promote a particular problem definition, causal interpretation, moral evaluation, and/or treatment recommendation'' \cite{entman1993framing}. By studying how misinformation is produced and how audiences believe it, as well as carefully surveying different means of news framing, we can have a strong motivation and guidance on how to address misinformation from a visual analytic perspective.

In this paper, we approach misinformation from three different perspectives. First, we discuss existing research on how misinformation is produced. More specifically, how news outlets and individuals use different methods to alter existing news or to fabricate completely new and false information. Next, we discuss what makes audiences believe and trust misinformation. Finally, we review over existing methods for identifying, detecting, and combating misinformation.

\subsection{Production of misinformation:} How and why misinformation is created is a complex topic, and studying it requires careful consideration of the type and intent of misinformation outlets, as well as different strategies they use to create and propagate ``effective'' articles. Several intents and reasons can be noted on why misinformation exists. These reasons and intents include economic (e.g., generating revenue through internet advertisements and clicks), ideological (e.g., partisan accounts focusing on ideology rather than truth), or political factors (e.g., government produced propaganda, meddling with other countries political processes) \cite{tambini2017fake,tandoc2017defining,arsenault2006conquering}. Driven by these intents, certain news outlets use different strategies to produce and diffuse misinformation. 

News media that produce misinformation often follow topics and agendas of larger mainstream partisan news media, but in some instances, they have the potential to set the agenda for mainstream news accounts \cite{vargo2017agenda}. Furthermore, they have a tendency to have narrow focuses on specific types of information that can manipulate specific populations. Often, we can observe extreme ideological biases towards specific political parties, figures, or ideologies \cite{spohr2017fake, arsenault2006conquering}. Some outlets take advantage of the psychological climate of audiences by focusing on fearful topics or inciting anger and outrage in audiences \cite{arsenault2006conquering}. Moreover, misinformation accounts tend to take advantage of cognitive framing of the audience by focusing on specific moral visions of certain groups of people adhere to \cite{entman2010media,arsenault2006conquering,lakoff2010moral}. They also filter information to focus on specific subsets of news, often focusing on specific political figures or places \cite{adams1986whose,allcott2017social}. 
 
Misinformation is not always present just in text. Biases and news framing are often presented in images rather than text to convey messages \cite{frenkel_2017}. The power of images is that they can contain implicit visual propositioning \cite{abraham2006framing}. Images have been used to convey racial stereotypes and are powerful tools to embed ideological messages \cite{messaris2001role}. Different means of visual and framing have been used to depict differences in political candidates from different parties \cite{grabe2009image}. For example, misinformation outlets use images by taking photos out of context, adding text and messages to the images, and altering them digitally \cite{mallonee_2017}. 
 
Another important aspect in the production and dissemination of misinformation, specifically in the age of social media, are social bots. Social bots are automatic accounts that impersonate real outlets and manipulate information in social media. They are sometimes harmless or useful, but are often created to deceive and manipulate users \cite{ferrara2016rise}. Social bots can like, share, connect, and produce misinformation \cite{lazer2018science}, they have the potential to amplify the spread of misinformation, and exploit audiences cognitive biases \cite{lazer2017combating}. It has been shown that bots are active in the early stages in the spread of misinformation, they target influential users and accounts through shares and likes, and might disguise their geographic locations \cite{shao2017spread}.

\subsection{Why and how we believe misinformation} Misinformation is falsified and fabricated information taking the form of real news. Equally as important to how misinformation outlets slant or fabricate news is produced, is to understand why and how people believe certain falsified information or take part in propagating the information. 
There are various social and psychological factors that effect the way we accept or reject information. It has been shown that collective means and social pressure affect our decision-making processes. Hence, we are likely to believe a news article to be true if our social group accepts it \cite{sloman2018knowledge, lazer2017combating, eckles2017bias}. Prior exposure to a statement also increases the likelihood that individuals believe it to be true \cite{pennycook2017prior}. These factors form echo chambers in our real and digital societies and effect our ability to judge information accurately. These echo chambers are also amplified by algorithmically controlled news outlets that expose users to news they are more likely to interact with \cite{sunstein2001echo, bakir2017fake, flaxman2016filter}. As individuals, we are also affected by confirmation bias and are more receptive of views that confirm our prior beliefs \cite{mynatt1977confirmation, lazer2017combating}. In contrast, the ability to think analytically has a positive effect on the ability to differentiate misinformation from real news \cite{pennycook2018cognitive}. The uncertainty of misinformation and the ability to qualitatively and quantitatively analyze misinformation also has some effect on individual's ability to distinguish misinformation \cite{karduni2018icwsm}.

\begin{figure*}[t]
  \centering
    \includegraphics[width=1.0\textwidth]{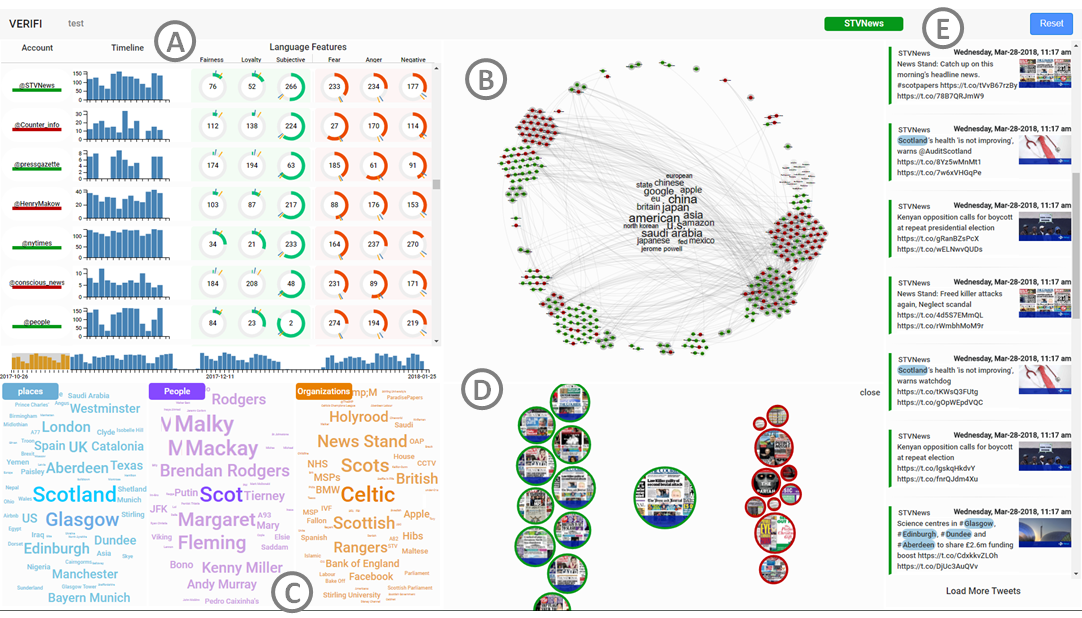}
  \caption{The Verifi2 interface is designed to support the investigation of misinformation sources on social media. The interface consists of 5 views: A) Account view, B) Social Network view, C) Entity Cloud view, D) Word/Image Comparison view and E) Tweet view.}
  \label{fig:interface}
\end{figure*}

\subsection{Battling misinformation} 
In order to combat the spread and influence of misinformation, we need to improve the structures and methods of curtailing misinformation to decrease audiences initial exposure, but we also need to empower individuals with the knowledge and tools to to better evaluate the veracity of the information \cite{lazer2018science}. 

The gravity of the issue has brought many researchers in computation to develop novel methods of detecting and identifying misinformation of various kinds. Many methods focus on detecting individual stories and articles. A number of fact-checking websites and organizations heavily rely on humans to detect misinformation \cite{snopes.com_2018,politifact}. Researchers have developed a model to detect click-baits by comparing the stance of a headline in comparison to the body \cite{bourgonje2017clickbait}. By focusing on sarcasm and satirical cues researchers were able to create a model that identifies satirical misinformation \cite{rubin2016fake}. Images have also been used to distinguish misinformation on Twitter. It has been shown that both meta user sharing patterns and automatic classification of images can be used to identify fake images \cite{gupta2013faking}. Using previously known rumors, classification models have been created to detect rumor propagation on social media in the early stages \cite{wu2017gleaning, shu2017fake}. Even though detecting misinformation on the early stages is extremely valuable, it has been noted that enabling individuals to verify the veracity and authenticity of sources might be more valuable \cite{lazer2018science}.

Social media data has been used in visualizations related to various domains including planning and policy, emergency response, and event exploration. \cite{karduni2017urban, maceachren2011geo, maceachren2011senseplace2} . Several visualization based applications have been developed that allow users to explore news on social media. Zubiaga et al. developed TweetGathering, a web-based dashboard designed for journalists to easily contextualize news-related tweets. TweetGathering automatically ranks newsworthy and trending tweets using confidence intervals derived from an SVM classifiers and uses entity extraction, several tweet features and statistics to allow users to curate and understand news on social media. \cite{zubiaga2013curating}.  Marcus et al. designed twitInfo, an application that combines streaming social media data several models including with sentiment analysis, automatic peak detection, and a geographic visualizations to allow users to study long-running events \cite{marcus2011twitinfo}. Diakopoulos and colleagues developed Seriously Rapid Source Review (SRSR) to assist journalists to find and evaluate sources of event-related news \cite{diakopoulos2012finding}. Using a classification model, SRSR categorizes sources to organizations, journalist/bloggers, and ordinary individuals. Furthermore, they use several features derived directly from Twitter as well as named entity extractions to allow journalists to understand the context and credibility of news sources. Even though these applications do not focus on misinformation, their lessons learned from dealing with social media are important inspirations for the design of Verifi.

There have been some efforts to classify and differentiate between verified and suspicious sources or to visualize specific types of misinformation such as rumors or click-baits. It has been shown that linguistic and stylistic features of news can be helpful in determining trustworthy news sources. Moreover, focus on different topics can be used to characterize news sources \cite{mukherjee2015leveraging}. On Twitter, social network relationships between news accounts, as well as emotions extracted from tweets have been useful in differentiating suspicious accounts from verified sources \cite{volkova2017separating}. Furthermore, using different language attributes such as sentiment, URLs, lexicon-based features, and N-grams models have been created that can accurately classify rumors \cite{lazer2017combating}. There are also a handful methods or systems that aim to enable humans to interact and understand veracity of sources or news articles. Besides the mentioned human curated fact-checking websites, There are systems that allow users to explore sources or (semi-) automatically detect news veracity on social media including FactWatcher \cite{hassan2014data}, Rumorlens \cite{resnick2014rumorlens}, and Hoaxy \cite{shao2016hoaxy}.  

Most of these systems allow some exploration and fact-checking on specific misinformation related factors. As a recent survey of more than 2,000 teachers regarding the impacts of technology on their students' research ability shows, current technologies should encourage individual to focus on a wide variety of factors regarding sources of misinformation. Inspired by previous research in psychology, social sciences, and computation; and In line with recommendations that focusing on empowering individuals and enabling them to differentiate sources are the best strategies to battle misinformation \cite{lazer2018science}, we developed Verifi2. Verifi2 employs an exploratory visual analytic approach and utilizes a combination of existing and new computational methods and is inspired by existing literature on misinformation, as well as a number of previous user studies on early prototypes to understand how users interact with misinformation \cite{karduni2018icwsm}.

\section{The Design of Verifi2}
In this section, we detail the inspirations for the Verifi2 system design, the tasks that Verifi2 aim to support, and distinguish Verifi2 from earlier prototypes that are specifically developed for user experiments to evaluate cognitive biases \cite{karduni2018icwsm}.

\subsection{Task Characterization}
\label{{sec:Tasks}}
Combating misinformation requires a multidisciplinary effort. On the one hand, the data mining and NLP research communities are producing new methods for detecting and modeling the spread of misinformation \cite{shu2017fake}. On the other hand, social scientists and psychologists are studying how misinformation is perceived and what intervention strategies might be effective in mitigating them\cite{PLOSONE2017}. More importantly, pioneers on misinformation research have been calling for an educational response to combating misinformation \cite{SHEG:2016, lazer2018science}. 
Verifi2 aims to contribute to this cause by bringing together findings from various disciplines including communications, journalism, and computation and raise awareness about the different ways news accounts try to affect audiences' judgment and reasoning. 
 In addition, following Lazer et al.'s recommendation \cite{lazer2018science}, Verifi2 aims to support the analysis of misinformation on the source/account level, in contrast to investigating one news story at a time.

Verifi2 is designed to support multiple tasks and enable users to evaluate the veracity of accounts at scale.
The tasks are partially inspired by our comprehensive review and categorization of the recent literature on misinformation mentioned in section 2. The list of task are also inspired by our discussions and collaborations with researchers that have been working on computationally modeling misinformation. The tasks include:

\begin{itemize}
    \item Task1: presenting and raising awareness of features that can computationally separate misinformation and real news;
    \begin{itemize}
      \item Task1A: presenting linguistic features of news accounts
      \item Task1B: presenting social network features of news accounts
  \end{itemize}
    \item Task2: enabling comparison between real and suspicious news sources;
      \begin{itemize}
      \item Task2A: presenting top mentioned entities to enable the comparison around an entity of interest
      by different news accounts
      \item Task2B: 
      enabling comparison of slants in usage of image postings and text between real and suspicious news accounts
    \end{itemize}
    \item Task3: supporting flexible analysis paths to help users reason about misinformation by providing multiple aspects that contribute to this complex issue.
\end{itemize}

The tasks drove the design of the Verifi2 system, including the back-end computation and the front-end visualizations. Details of the Verifi2 system are provided in section \ref{sec:Verifi}.

\begin{figure*}[t]
  \centering
    \includegraphics[width=1.0\textwidth]{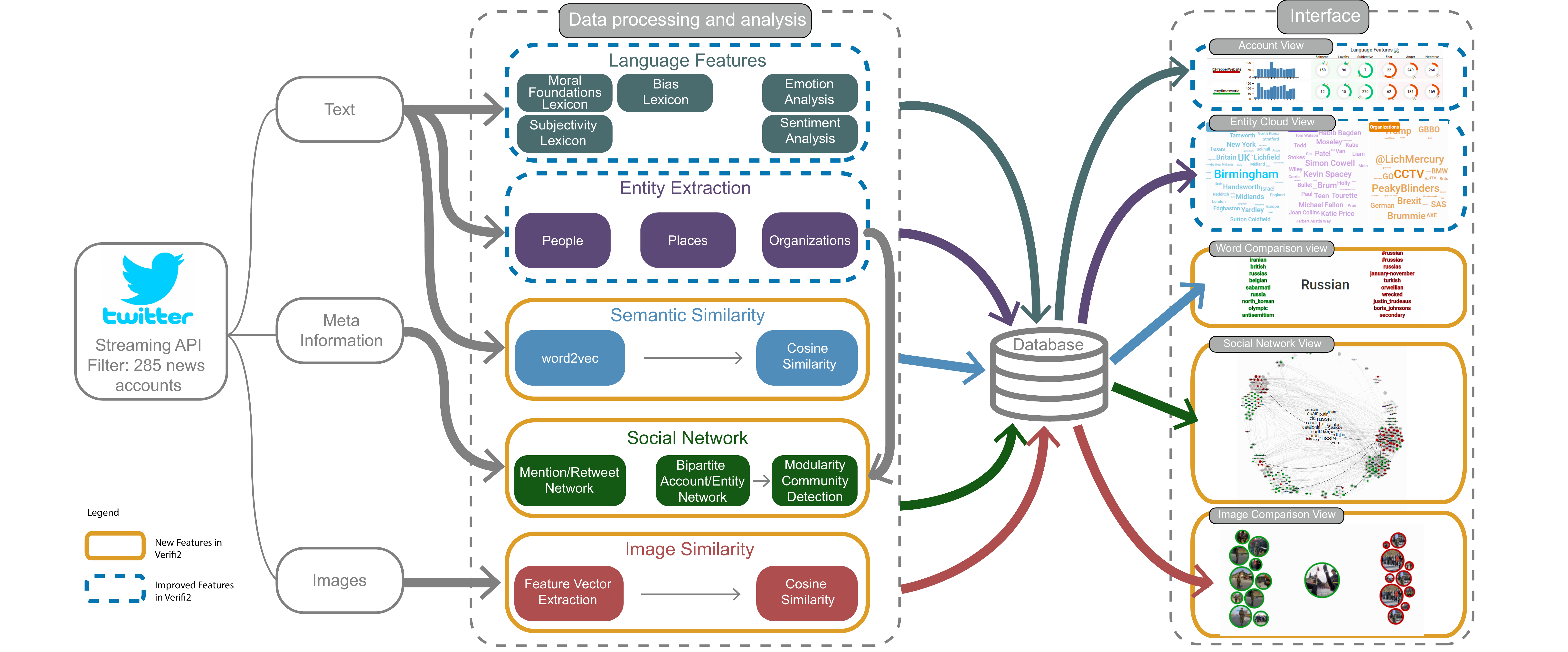}
  \caption{Verifi2 pipeline. Tweets, meta information, and images from 285 accounts are collected from Twitter streaming API. Language features, named entities, and word embeddings are extracted from the tweets. A social network is built based on the mention/retweet relationships. The accounts are clustered together using community detection of a bipartite account/named entity network. Image similarities are calculated based on feature vectors. All of these analysis are stored in a database and used for the visual interface.}
  \label{fig:pipeline}
\end{figure*}

\section{Verifi2}
\label{sec:Verifi}
In this section, we provide detailed descriptions on the computational methods, the Twitter dataset we collected for misinformation investigation, and the visual interface. The overall system pipeline of Verifi2 is shown in Fig. \ref{fig:pipeline}. 

\subsection{History of Verifi} 

Verifi2 is a product of incorporating feedback from 150+ users from two experiments on decision-making on misinformation 
\cite{karduni2018icwsm} and a comprehensive literature review on how misinformation is produced, believed, and battled. 
Verifi1\footnote{https://verifi.herokuapp.com/} was is an interface that is built for conducting studies on cognitive biases using decision-making tasks around misinformation. It contains a small number of masked accounts and simple visualizations. Users would use Verifi1 to make decisions on whether an unknown account is real or fake based on given cues and information \cite{karduni2018icwsm}. After conducting a thorough literature review and further analyzing the user studies, we realized that users are will not always have a visual analytics system at hand to make real-world decisions. Therefore, we developed Verifi2 with a new goal: Enabling users to explore news through a visual analytic system, learn about how news sources induce bias and slants in their stories, and ultimately transfer this knowledge in their real-world scenarios. However, The experiments and studies conducted using Verifi1 highlighted important points that guide the new features and goals of Verifi2:
\begin{itemize}
    \item Users rely heavily on their qualitative analysis of news article in conjunction of provided information. This result guided us to create completely new features that support qualitative analysis of news/context of news such as usage of visual imagery, topic similarity, and social network communities.
    \item Verifi1 included a simple force directed visualization of an undirected mention/retweet network. Even though these networks were shown to be computationally discriminating, they proved to be extremely misleading for some specific cases. In Verifi2 we provide a new network visualization with directed edges and topic communities.
    \item Verifi1 included a view for map visualization of place entities. This view was the least used view in Verifi1 and users suggested it was difficult to use. In Verifi2 we replace that view with a word-cloud visualization of the place entities.
    \item Verifi1 included linguistic features that proved to be vague or hard to interpret for users. We removed those features including non-neutral language, added new color coding metaphors and encodings for medians and averages of each score.
\end{itemize}
Finally, even though the design language of Verifi2 is influenced by Verifi1, Verifi2 incorporates new features or important improvements in every step within the pipeline (Figure \ref{fig:pipeline}), including more social media news accounts, different language features, new algorithms to support the comparison between real and suspicious accounts, and modified visual representations compared to Verifi1.

\subsection{Data}
Verifi2 is designed to enable investigation of misinformation on a source-level. Therefore, our data collection starts with identifying a list of verified and suspicious news accounts. The list was obtained by cross-checking three independent third-party sources and the list has been used in prior research on identifying features that can predict misinformation \cite{volkova2017separating}. The list includes 285 news accounts (166 real and 119 suspicious news accounts) from around the world. 
We then collected tweets from these accounts between October 25$^{th}$ 2017 to January 25$^{th}$ 2018 from the public Twitter Streaming API. As a result, more than 1 million tweets were included in our analysis. In addition to the account meta data and the tweet content, some tweets include links to images. Since images can play an important role in making a tweet more attractive and thus receiving more attention \cite{messaris2001role}, we further collected all images for analysis. As a result, all tweets, account meta data, and images were stored in our database and were accessed using NodeJS. The tweets were then processed through a variety of different computational methods to support the analysis and comparison between sources of real news and misinformation.

\subsection{Data Analysis and Computational Methods}
\label{sec:Task}

This section will describe the data analysis and computational methods used to enable users to achieve the tasks Verifi2 aims to support.

\subsubsection{Data analysis to support Task 1} 
\label{sec:task1}

\textbf{\newline Supporting Task 1A\textendash Language use by news accounts:} Research on misinformation and framing suggests that suspicious accounts are likely to utilize anger or outrage-evoking language \cite{arsenault2006conquering,lakoff2010moral,spohr2017fake}. Verifi2 enables users to explore language use of news accounts. To achieve this task, we characterize language use by utilizing a series of dictionaries developed by linguists and a classification model for predicting emotions and sentiment. We started with a broad set of language features, including moral foundations \cite{graham2009liberals,haidt2007morality}, subjectivity \cite{wilson2005recognizing}, and bias language \cite{recasens2013linguistic}, emotion \cite{volkova2015inferring}, and sentiment \cite{liu2005opinion} (Fig. \ref{fig:pipeline} Language Features). Each of these features consists of various sub-features/dimensions. For example, the moral foundations includes five dimensions, namely fairness/cheating, loyalty/betrayal, care/harm,  authority/subversion, and purity/degradation. We extract six different emotions (fear, anger, disgust, sadness, joy, and surprise) and three sentiment dimensions (positive, negative, and neutral). Overall, we collected over 30 dimensions that can be used to characterize language use. In \cite{karduni2018icwsm}, we performed a random forest to rank the dimensions based on their predictive power for a news account being real or suspicious.  Among all of the calculated features, high score in fairness, loyalty, subjectivity was found to be predictive of real news accounts and fear, anger, and negative sentiment were found to be predictive of suspicious news accounts. The details for all language features can be found at \cite{karduni2018icwsm}, which presented a study leveraging Verifi1. For Verifi2, we re-examined the language features and removed the ones participants found confusing. As a result, six language features were included in the Verifi2 interface.  

\textbf{Supporting Task 1B\textendash Understanding relationships between news accounts:} There are multiple ways to investigate the relationships among real and suspicious news accounts on Twitter. On the one hand, previous research \cite{volkova2017separating} suggests that it is more likely for suspicious news accounts to mention or retweet real news accounts but not vice versa. On the other hand, literature on misinformation and framing shows that accounts have the tendency to focus on a subset (narrow set) of news topics compared to real news accounts \cite{spohr2017fake,arsenault2006conquering}. To present these two important aspects, we constructed a social network of all 285 news accounts based on both the retweet/mention relationship and the common entities they mention. 
To model the relationship among accounts based on their entity mentions, we constructed a bipartite network with two types of nodes: news accounts and entity mentions (people, places, and organizations). If an account mentions an entity there would be an edge between the two, weighted by the number of times the account mentioning that entity. We then applied maximum modularity community detection \cite{Newman8577} on the bipartite network to identify communities of news accounts. 

The community detection result contains 28 communities each with a number of accounts and entities with meaningful relationships. To briefly illustrate the community detection results, we describe three salient communities. The largest one contains 49 accounts (38 suspicious and 11 real). The top entities in this community include Russia, ISIS Iraq, Catalonia, Putin, and Catalonia which suggest an international focus of this community. Indeed, after examining self described locations of the accounts from their Twitter profiles, a vast majority were from Europe, Russia, and the Middle East. The second largest community contains 41 nodes (31 real and 10 suspicious), with top entities in the community including Americans, Congress, Democrats, Republicans, and Manhattan
which suggest a focus on American politics. Investigating the locations of these news accounts show that almost all of these accounts are from the United States. Another large community comprised entirely of suspicious accounts mention entities including Muslims, Islam, Lee Harvey Oswald, and Las Vegas which hints towards sensational topics related mostly to the United States. 

\subsubsection{Data analysis to support Task 2} 
\label{sec:task2}
\textbf{\newline NLP methods to support Task 2A\textendash Understanding the top mentioned entities but real and suspicious news accounts:} 
Extant research shows that different accounts tend to have specific focus on topics such as places and political figures. By highlighting these entities from the text of tweets, we allow users to explore topical features of these news accounts. By extracting three types of named entities (people, places, and organizations) we can enable users to both easily explore topics of interest and compare how different accounts treat specific entities or topics. To extract named entities, we used python SpaCy.\footnote{https://spacy.io/} After semi-automated clean up of the false positives and negatives, we saved the results in the database along with the tweets to be utilized in the interface.

Based on the top entities, Verifi2 further integrates computational methods to support comparing real and suspicious account postings around the same issue.
Being able to compare the keywords usage around the same issue between real and suspicious news accounts contributes to the understanding of how different news media outlets frame their news. To this aim, we integrated a word embedding model \cite{NIPS2013} in Verifi2 to present the most semantically related words to entities selected interactively by an user. We first obtained vector representations of all words in our tweet collection using TensorFlow\cite{tensorflow}. We then use the selected entity as the query word and identify the top keywords by calculating the cosine similarity between the query word and all words in the vocabulary. We performed this process separately for real news and suspicious news accounts. As a result, we were able to identify and display the most semantically similar keywords to a query word from both real news and suspicious news accounts' postings.

\textbf{Computational methods to support task 2B\textendash Compare real and suspicious account preference on images: }
Research on misinformation and news framing shows that visual information including images are used heavily to convey different biased messages without explicitly mentioning in text \cite{abraham2006framing,messaris2001role,grabe2009image}. To this aim, we utilized images from our 285 tweet accounts to enable users to compare and contrast the images posted by real and suspicious news accounts. We use the ResNet-18 model with pre-trained weights to extract a feature vector of 512 dimensions for every image from the ``avgpool'' layer\footnote{https://tinyurl.com/y9h7hex} \cite{he2016deep}. For any image that a user interactively selects as the query, we identify the top 10 similar images from both real news and suspicious news accounts measured by cosine similarity. Therefore, users can compare the images with accompanying tweets from real and suspicious accounts given an input image. 

\subsection{Visual Interface}

Verifi2 includes a visual interface with multiple coordinate views to support a variety of analytical tasks regarding misinformation described in section \ref{sec:Task}. The interactions are designed to support \textbf{Task 3} in that it enables flexible analysis strategies to make sense of the veracity of news accounts. The visual interface consists of five views: Account view, Social Network view, Entity Cloud view, Word/Image Embedding view and Tweet view. Through a series of interactions, each view enables users to focus specific aspects of misinformation while the rest of the views update accordingly. 
In this section, we will describe the features and affordances of each view while detailing how they are coordinated via users interactions. 

\subsubsection{Account View} 
The Account view shows tweet timeline and language features of 285 Twitter news accounts (Fig. \ref{fig:interface}A 
). Real news accounts are underlined by green lines and suspicious news accounts are underlined by red lines 
. The bar chart shows the daily tweet count of the account of the selected time range 
The six language features (fairness, loyalty, subjectivity, fear, anger and negativity) introduced in section \ref{sec:task1} are represented by donut charts (scaled from 0-100). The numbers in the donut charts indicate the ranking of each language feature per account based on its score. For example, as seen in Fig. 
\ref{fig:interface}A the account @NYTimes ranks high on fairness (34$^{th}$ out of 285) and loyalty (21$^{th}$) while ranking low on anger (237$^{th}$) and negativity (270$^{th}$). Two lines in the donut chart indicate mean (orange) and median (blue) of the language feature to facilitate the understanding of how a particular account score given the mean and median of all accounts. The language features are displayed in two groups based on the correlations with the news accounts being real or suspicious. Overall, real news accounts have positive correlation with fairness, loyalty, and subjectivity while suspicious news accounts have positive correlation with fear, anger, and negativity. This view is connected to the analysis described in section \ref{sec:task1} and supports Task1A. 

Clicking on an account name filters the data for all views to only tweets from the selected account. Hovering the mouse over the name of each account invokes a tooltip showing the description of the hovered news account based on its Twitter profile. Users can sort the accounts based on any language features and explore how real and suspicious news accounts differ in language use.

\subsubsection{Social Network View}

Based on the analysis introduced in section \ref{sec:task1}, the Social Network view present information on both retweet/mentions between news accounts, and the news account communities constructed based on the entities they mention (Fig. \ref{fig:interface}B). The visualization adopts a circular layout. Each node in the view represents a news account, colored in either red (suspicious) or green (real). The nodes are positioned at the perimeter of the circle. The directed links between accounts are established by retweet and mention, while the account communities are determined based on entity mention. To avoid adding more colors to denote communities, we alternated a light and darker shades of gray in the nodes background for adjacent communities to present the clusters. This view supports Task1B. 


Hovering on a node highlights all connections of the node to other nodes by with outgoing and incoming edges. An incoming edge represents the account is being mentioned or retweeted while an outcoming edge represents the account retweeted or mentioned another account. Moreover, hovering on a node shows a word cloud of named entities related to its community. 
Furthermore, users can pan and zoom in the Social Network view to switch focus between an overview of the social network and particular communities for analysis. 

\subsubsection{Entity Cloud View}
The Entity Cloud view presents three word cloud visualizations for designed for people, places and organizations respectively (Fig. \ref{fig:interface}C). Users can get a quick overview of the top mentioned named entities by all or selected accounts. We applied three colors - blue, purple and orange for each word cloud to distinguish mentions of these entities. The same color assignment is used in the Tweet view to highlight mentions of different entities in individual tweets. The Entity Cloud view supports Task2A. 

Clicking on each entity in any of these world clouds filters the data to include tweets that contain these entities. It is important to note that the filters in Verifi stack, in that if an account is selected, then clicking on an entity shows tweets from that account containing that entity. Moreover, clicking on entities opens up the Word Comparison view to enable comparison of related terms from real and suspicious sources.

\begin{figure}[t]
  \centering
    \includegraphics[width=0.9\columnwidth]{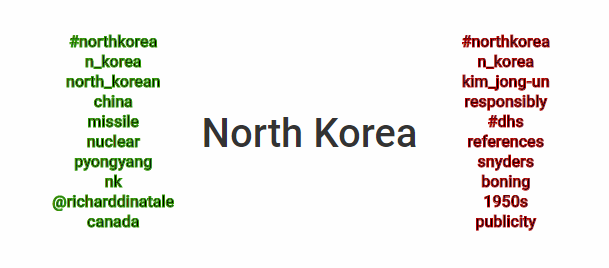}
  \caption{Word Comparison view. The query word is ``North Korea'', the top related keywords from real news accounts are shown on the left while the keywords from suspicious accounts are displayed on the right.}
  \label{fig:word-compare}
\end{figure}

\subsubsection{Word/Image Comparison View}
The Word Comparison view (Fig. \ref{fig:word-compare}) is visible when users clicks on an entity from the Entity Cloud view. The view allows users to compare top 10 semantically similar words from real (Fig. \ref{fig:word-compare} left) and suspicious news accounts (Fig. \ref{fig:word-compare}) right) to the selected entity. The user can find common and distinct words from either news sources. The user can further the comparison task by clicking on one of the semantic words. This action filters the tweets to those that include both selected entity and similar word. This view supports Task2A. 

Enabled by the analysis described in section \ref{sec:task2}, the Image Comparison view (Fig. \ref{fig:interface}D) displays the top 10 similar images from real and suspicious news accounts given a query image. The central image circle of this view is the query image. To the left and right of this image are the top similar images from real and suspicious news accounts. The size of the circle encodes the cosine similarity value; with larger circles capturing the images being more similar. Hovering on an image shows the enlarged image as well as the associated tweet, account, and cosine similarity to the selected image. Using this circular encoding, users can easily see whether images most similar to a selected image or mostly from real news accounts or suspicious accounts. This view supports Task2B. 

\subsubsection{Tweet View}
The Tweet view (Fig. \ref{fig:interface}E) provides details and allows users to read the actual tweets. This view enhances the reading experience by highlighting mentioned entities with the same color code as in the Entity Cloud view. The Tweet view also shows thumbnail images if a tweet links to an image. Hovering on an image thumbnail enlarges the hovered image for better legibility. Clicking on an image of interest opens the Image Comparison view to enable comparison of most similar images from real and suspicious news accounts.

\section{Usage Scenario:  exploring news related to an organization}


\begin{figure}[t]
  \centering
    \includegraphics[width=1.0\columnwidth]{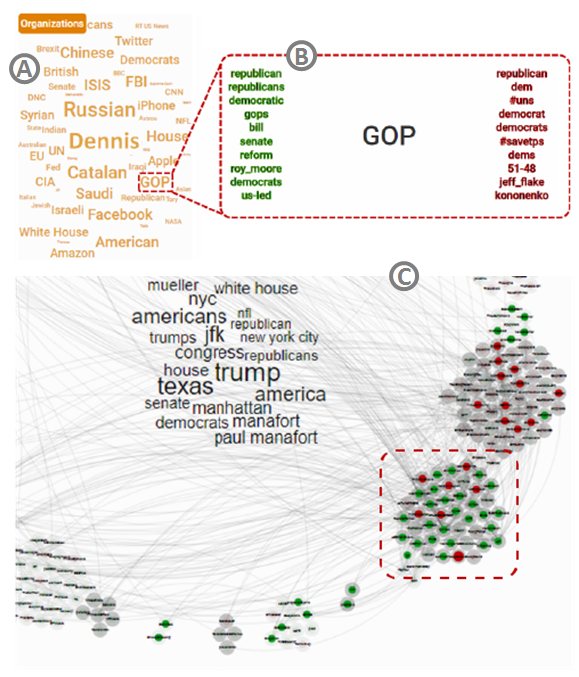}
  \caption{Top: Illustrating comparison between top semantically related words to the entity ``GOP'' (The Republican Party). Bottom: Community of news accounts in the social network that most mention the term GOP.}
  \label{fig:case1}
\end{figure}

\begin{figure}[t]
  \centering
    \includegraphics[width=1.0\columnwidth]{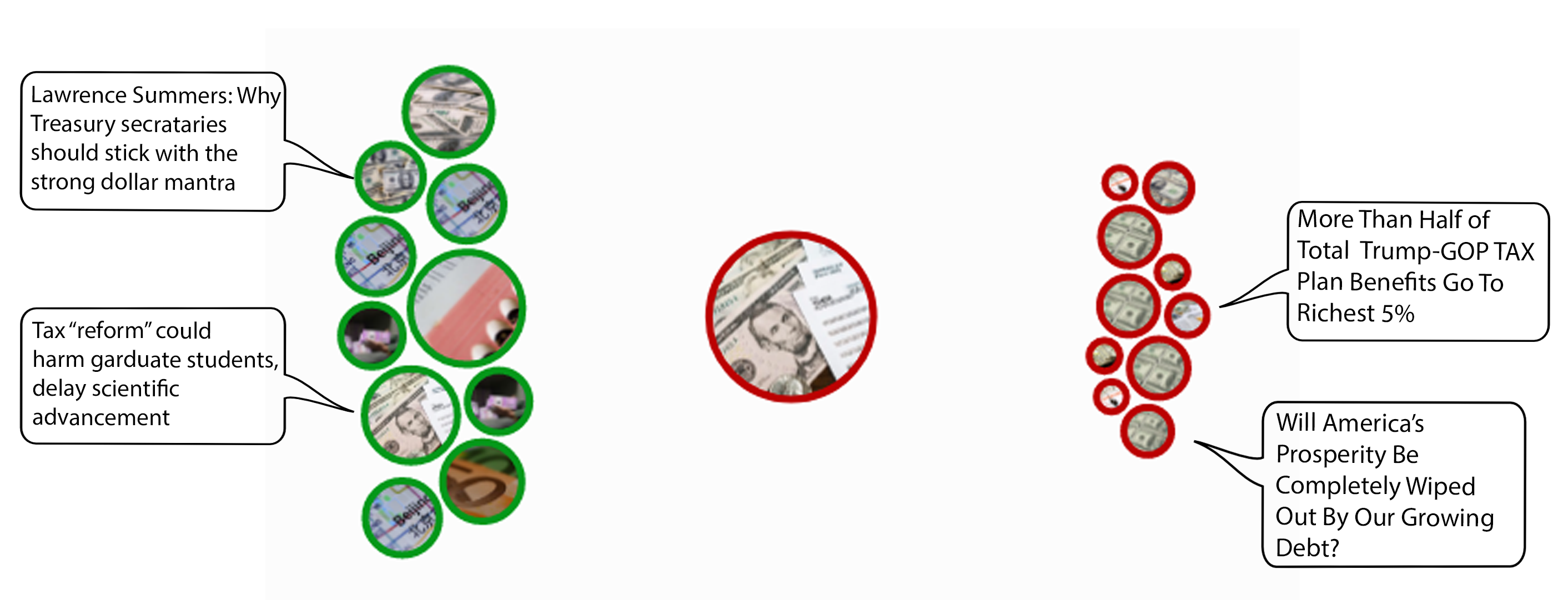}
  \caption{comparison of images/tweet pairs between real and suspicious news shows how these groups use images differently. 
  }
  \label{fig:scenario1}
\end{figure}

\begin{figure*}[t]
  \centering
    \includegraphics[width=1.0\textwidth]{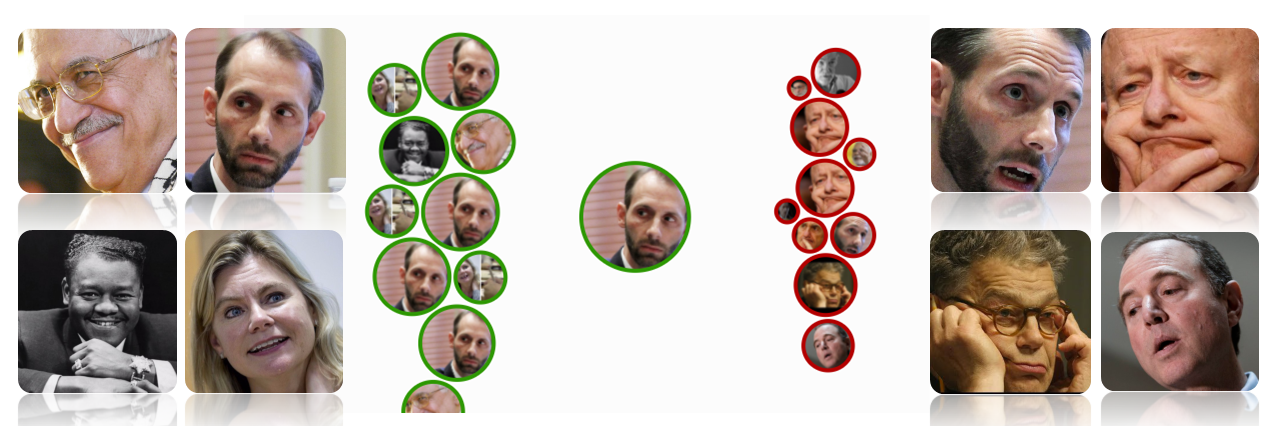}
  \caption{The Image Comparison view highlights how suspicious accounts frame images to convey messages.}
  \label{fig:facialcoding}
\end{figure*}

In this scenario, the user is interested in exploring the news regarding the political climate in the United States. After launching Verifi2, The user starts by looking at entities in the Entity Cloud View (Fig. \ref{fig:case1}A). She immediately observes that one of the top mentioned organizations is GOP. By clicking the word in the Entity Cloud view, the interface updates to show tweets that mention GOP in the Tweet view. At the same time, the Word Comparison view gets activated so she can compare words most related to GOP from either real or suspicious news accounts (Fig. \ref{fig:case1}B). She observes that some of the closest words from real news (green colored words) are republicans, democratic, gops, bill, senate, and reform 
mostly ``neutral'' words that are directly related to the organization. She also observes that the related words from suspicious accounts (red colored words) are different. They include republican, dem, \#UNS, democrat, \#savetps, and jeff\_flake.

Through comparing the top related words to ``GOP'', she finds the words dem, democrat, and democrats that are not among the top 10 words on the real news list. By clicking on the word ''dem'', she can now cross-filter with the word ``GOP'' and ``dems'' and see how different accounts report their stories involving the two keywords. She finds out that the majority of the tweets that include both keywords in the Tweet view are from suspicious accounts. One of the tweets is a Russian account retweeting an Iranian news account regarding a weapons ban proposal: 
\textit{``GOP fails to back Dems' weapons ban proposal''}. Other tweet looks extremely sensational and opinionated: \textit{``BREAKING: Dems and GOP Traitors Unite Against Roy Moore''} and \textit{``Enraged Liberals Snap After Senate Dems Get Treated Like Rag Dolls, Cave to GOP to End Govt...''}

The user then clicks on the word ``bill'' from the real news list which is not found on the misinformation list. Most of the tweets seems to be reporting events with neutral language about different bills:
\textit{``Rauner calls transparency bill 'political manipulation,' angering GOP''}, \textit{``Property-tax deduction could help GOP reform bill''} and \textit{``GOP leaders plan to hold votes on the bill early next week, first in the Senate and then the House.''}

Having learned from this comparison that the way suspicious news accounts frame news about GOP might be different that real news accounts, she continues to explore differences between accounts in the Social Network view. The view highlights the accounts that have used the term GOP. She notices that most of the accounts are concentrated in one community (See Fig. \ref{fig:case1}C). The top mentioned entities by this cluster shown in the graph include republicans, congress, house, and senate which are concepts related to American politics. After zooming into the cluster, she realizes that even though most of the accounts in this cluster are real accounts, there are a number of suspicious news accounts (in red). She then clicks on one of the red accounts to investigate how that accounts treats the GOP organization.
 She then goes to the Tweet view to examine tweets from these suspicious accounts. The first tweet is \textit{``Democrats Were Fighting for the Rich in Opposing the GOP's 401(k) Cut Proposal''}. The tweet includes an image that shows dollar bills and tax papers. By clicking on the image, she is able to compare the related images from real and suspicious news accounts (Fig. \ref{fig:scenario1}). She observes that all the images include pictures of money. She notices that one real news account is using the same image. By hovering on the images, she compares the tweets that link to these images from real and suspicious account. One tweet from a real news account is about tax reform: \textit{``Tax `reform' could harm graduate students, delay scientific advancement...''}. She then finds another image again with dollar bills and tax papers from a suspicious account: \textit{``More Than Half of Total Trump-GOP Tax Plan Benefits Go to Richest 5\%''}. She observes that this tweet has a suggestive tone in comparison to the more neutral tone of the tweet containing the similar images from real news.

\section{Expert Interviews} 
Our goal with Verifi2 is to help individuals learn about the different ways suspicious news accounts are showed to introduce slants, biases, and falseness into their stories. Moreover, we aim to produce a system to create help users make decisions in real-world scenarios. To assess our system on these grounds, 
we conducted five semi-structured interview sessions with experts from different disciplines who work on battling misinformation or conduct research on misinformation. These five individuals were university professors from education/library sciences, communications, digital media, psychology, and political science. Naturally, each of our interviewees had a different approach to understanding and battling misinformation which resulted in valuable commentary on both the features of Verifi2, as well as the potentials for it to be used in real-world attempts to battle misinformation. The interviews were recorded, transcribed, and analyzed.

\subsection{Interview Setup}
Interview sessions took between 45-60 minutes. Before showing the interface to the interviewees, we asked them to introduce themselves, their professions, and their specific concern with misinformation. Next, we asked general questions about misinformation and how technology can assist in attempts to battle it. We also inquired our interviewees about how they differentiate between misinformation and real information, suspicious and real news sources, and the reasons audiences chose to believe misinformation. After the initial questions, we opened Verifi2 and started thoroughly explaining the different parts of the interface. The interviewees made comments and asked questions regarding the features throughout the interface walk through. Finally, we asked to important closing questions. Namely if they thought of other features that needs to be added to the interface, and what potential uses they see for Verifi2. In this section, we will summarize the important lessons learned from these interviews. Direct quotes from interviewees are slightly edited for legibility.

\subsection{Facial action coding: A psychologist's perspective}

Biases and slants can be hidden not only in text, but also in the images. These hidden farmings can activate our mental frames and biases without us noticing them \cite{grabe2009image, messaris2001role}. During our interviews with experts conducting research about misinformation, a professor in psychology who specializes in emotion analysis made observations about the differences between real and suspicious accounts on their usage of graphics/images. After receiving an introduction of the interface, the psychologist started exploring the image comparison feature in Verifi2. She was particularly interested in close-up images of individuals. Given her expertise in emotion contagion, her hypothesis was that real vs. suspicious news outlets may frame the visual information in tweets using subtle emotional encoding. She focused on the facial expressions in the images used in suspicious accounts compared to real news accounts. 

Clicking on the picture of a person active in politics and examining the different top related images from real or suspicious accounts, she said: \textit{``These [pointing at the images from suspicious sources] are all more unattractive. you see this with [polarizing political figures], all the time whereas these [pointing at images from real news accounts] are not, they are all showing positive images, [\ldots]. So these people [real news images] look, composed! these [suspicious news images] less-so. So fake news apparently is also much more likely to [use negative imagery].[\ldots] That's exactly what I do. When I started I did facial action coding. So you can actually do individual coding of facial muscles and they correspond with particular emotions and that is also something that you can [use the interface for]''}. 
Fig. \ref{fig:facialcoding} shows the pictures this expert was using as an example to show the way suspicious accounts frame their images using emotional encoding.

\subsection{
A spectrum of trustworthiness rather than binary classes}

All of our interviewees acknowledged the challenge in defining misinformation. As well as the complexity of the issue which involves both news outlets with different intents, as well as audiences with different biases. Their definitions of misinformation were different but with some common features. They made notes on our binary choice (real vs. suspicious) and how it is important to communicate the complexity of misinformation and move away from a simple binary definition thus not alienating the people who might be most prone to the dangers of misinformation.



The expert in digital media discussed the subtlety of the 
of the misinformation problem and hinted towards revisiting how we classify news accounts \textit{you have maybe a factual article, but the headline is focusing on a particular part of that article and a kind of inflammatory way that actually has nothing to do with the constitution}. Our communications expert mentioned that misinformation could be in forms other than news: \textit{: ``Misinformation could be in different settings, not just narrowly focused on news in particular, however, [completely false news] is probably one of the most problematic forms of misinformation because it has the potential to reach audiences''}. 



Two of the interviewees mentioned that the categorization of the real vs. suspicious accounts should be within a spectrum rather than a binary decision. Our digital media expert elaborated on this point: \textit{``there's a couple of websites or Twitter feeds that I've been following that are really, problematic because they're basically funded by NATO and they're there to track Russian misinformation and propaganda. But the really interesting thing about it is that they're using language that's very propagandistic themselves.''} She then continued about how the interface could respond to these gray areas: \textit{``maybe rather than ascribing them [the accounts] as being real or [suspicious], they can be scored on a continuous spectrum taking into accounts all features in Verifi."}

\subsection{Verifi2 as a potential tool for education on misinformation}
Two out of five interviewees have been working on the ``Digital Polarization'' project, which is part of a national program to combat fake news \footnote{http://www.aascu.org/AcademicAffairs/ADP/DigiPo/}. Therefore, educating college students about misinformation has been on the top of their minds. After walking he interviewees through the usage of Verifi2, each had different thoughts on how the interface would be useful within their tasks.

The digital media professor discussed how Verifi2 would be a useful tool for teaching students about how to compare different sources. However, she was concerned about alienating individuals with different political preferences: 
\textit{``\ldots someone may subscribe to some fake news websites, and that's what they think [is true]. And I think that's all true if you presented this to them and then they saw that their preferred website came up as as fake information. Um, what happens is that'll just set their back up against a wall and they'll be like, well, no, this is totally wrong.''}. she then continued to describe how the interface would be useful for high school and college students if it allowed them to come to conclusions themselves: \textit{``I think it would be such a great tool for teaching people how to evaluate information. I'm especially thinking of high schoolers [\ldots] I think about some of my freshman students coming in and they're really good at parsing out information that's presented visually and they would be really good at navigating something like this and say, oh, OK, this is leaning towards true or this is leaning towards fake..''}

The political science professor emphasized the importance of creating a tool which discourages individuals from partisan motivated reasoning: \textit{``one of the most important things that your tool would have to do is to help people disengage from a partisan motivated reasoning or predisposition based reasoning, which is actually really hard to do. I mean, you can cue people and try to reduce it, But we have an inherent tendency to shore up our existing beliefs, but a tool that, that somehow  gets people to think about things in atypical ways [would be really beneficial]''}

There was also a discussion of using Verifi2 as an assignment for students through which they would apply their knowledge in media literacy to decide whether accounts are trustworthy or not. The education expert described this and how Verifi2 can be applied in a number of different educational settings: \textit{``
 I'm thinking for a student assignment, we could [for example] pick 5 reliable sources and 5 unreliable ones. [\ldots] I could even see this potentially being useful not just in these targeted classes [that we teach] but in some of the liberal studies courses, that are more Gen ed requirements. I can see this, depending on the right instructor, if we knew about this tool that could educate people about [misinformation], they could be developing assignments for their writing classes or inquiry assignments. Overall, I see a lot of potential.''}

Our expert interviews highlight the potentials for Verifi2 to be used in scenarios where experts educate individuals about differences between real and suspicious sources of news. However, they highlighted the delicacy required to take on the complicated task of educating younger individuals on these concepts. Their recommendations included simpler interface and allowing users to decide the trustworthiness of the accounts based on their own assessment.

\section{Discussion}
The final design of Verifi2 is inspired by three threads of research and experiments: data mining and NLP, social sciences, and direct user feedback from prior research on decision-making on misinformation \cite{karduni2018icwsm}. 
With this in mind, Verifi2 supports calls to address misinformation from a multidisciplinary approach \cite{lazer2018science}.
To evaluate our system, we interviewed experts from multiple disciplines who focus on the problem of misinformation, each who provided different feedback. First, when we are dealing with the ever-increasing but complex topic of misinformation, we need to pay attention to the different types of misinformation as well as the intents of such producers of misinformation. Additional consideration must be made as well on the vulnerabilities and biases of consumers of information. To better deal with this issue, our interviewees recommended allowing users to 
make modifications on the systems encoding on suspicious or real accounts based on the many qualitative and quantitative evidence available through the system. In addition, they emphasized the importance of moving from a binary representation of accounts (real vs. suspicious) to a spectrum. One solution to both these suggestions is to enable human annotations (e.g., adjust binary classifications). By allowing users to annotate, we will increase user trust. Human annotations will be critical for accurate and updated results. Our current data is based on third-party lists that, while are good proxies, may change over time. Human annotation could become critical to validate, filter, and override. Combined with Visual interactive labeling (VIL) like CHISSL \cite{arendt2018chissl}, human annotators could provide real-time feedback on the accounts (e.g., cognitive psychologists who label extreme emotions).

Like many other researchers, our interviewees struggle with how to educate individuals on the spread of misinformation. They mentioned that Verifi would be a very useful tool to enhance traditional strategies such as media literacy courses. However, their responses made apparent the fact that the task of educating individuals requires much more work and rigor. It was mentioned that some simplifications and proper training needs to be done in order to make Verifi2 ready for education, especially for younger audiences. We are working with the researchers to modify, clarify, and prepare Verifi2 for such educational applications. Moreover, different educational settings require slightly different modifications. These modifications include enabling Verifi2 to focus only on a subset of the accounts or to enable users to observe completely new accounts, creating an infrastructure for Verifi2 to be used as course assignments, and partnering with educational institutions to deploy Verifi2.

\section{Conclusion}

While not a new phenomenon, misinformation as an application of study is in its infancy \cite{lazer2017combating}. In such an environment, visual analytics can play a critical role in connecting across disciplines \cite{lazer2018science}. In this paper, we introduced the Verifi2 system that allows users to approach misinformation through text, social network, images, and language features. We described case studies to show some of the ways users can utilize Verifi2 to understand sources of misinformation. Finally, we interviewed a diverse set of experts who commonly focused on misinformation, but had different approaches. We learned that Verifi2 can be highly useful to be deployed as an educational tool. But most importantly, we learned that Visual Analytics can be an extremely useful tool to battle misinformation. This, however, requires more work and a common goal from the visualization community to address the complexities of such issues. We call for visualization researchers to take on this task, to make our democracies more resilient to the harms of misinformation.

\section*{Acknowledgements}
The research described in this paper was conducted under the Laboratory Directed Research and Development Program at Pacific Northwest National Laboratory, a multiprogram national laboratory operated by Battelle for the U.S. Department of Energy. The U.S. Government is authorized to reproduce and distribute reprints for Governmental purposes notwithstanding any copyright annotation thereon. The views and conclusions contained herein are those of the authors and should not be interpreted as necessarily representing the official policies or endorsements, either expressed or implied, of the U.S. Government.
\bibliographystyle{ACM-Reference-Format}
\bibliography{VerifiSystemPaper}
\end{document}